\documentclass{Interspeech2024}
\usepackage{xcolor}
\usepackage{multirow}
\usepackage{cite}

\interspeechcameraready

\title{CtrSVDD: A Benchmark Dataset and Baseline Analysis for Controlled Singing Voice Deepfake Detection}

\name[affiliation={1}]{Yongyi}{Zang}
\name[affiliation={2}]{Jiatong}{Shi}
\name[affiliation={1}]{You}{Zhang}
\name[affiliation={3}]{Ryuichi}{Yamamoto}
\name[affiliation={2}]{Jionghao}{Han}
\name[affiliation={4}]{\\Yuxun}{Tang}
\name[affiliation={5}]{Shengyuan}{Xu}
\name[affiliation={5}]{Wenxiao}{Zhao}
\name[affiliation={5}]{Jing}{Guo}
\name[affiliation={3}]{Tomoki}{Toda}
\name[affiliation={1}]{Zhiyao}{Duan}

\address{
  $^1$University of Rochester, Rochester, NY, USA \
  $^2$Carnegie Mellon University, Pittsburgh, PA, USA \\
  $^3$Nagoya University, Nagoya, Japan \
  $^4$Renmin University of China \
  $^5$Timedomain.ai, Beijing, China
}
\email{svddchallenge@gmail.com, you.zhang@rochester.edu}

\keywords{singing voice deepfake detection, anti-spoofing, benchmark, dataset}

\begin{document}

\maketitle

\begin{abstract}
Recent singing voice synthesis and conversion advancements necessitate robust singing voice deepfake detection (SVDD) models. Current SVDD datasets face challenges due to limited controllability, diversity in deepfake methods, and licensing restrictions. Addressing these gaps, we introduce CtrSVDD, a large-scale, diverse collection of bonafide and deepfake singing vocals. These vocals are synthesized using state-of-the-art methods from publicly accessible singing voice datasets. CtrSVDD includes 47.64 hours of bonafide and 260.34 hours of deepfake singing vocals, spanning 14 deepfake methods and involving 164 singer identities. We also present a baseline system with flexible front-end features, evaluated against a structured train/dev/eval split. The experiments show the importance of feature selection and highlight a need for generalization towards deepfake methods that deviate further from training distribution. The CtrSVDD dataset\footnote{\url{https://zenodo.org/records/10467648} (Train/Dev)}\footnote{\url{https://zenodo.org/records/10742049} (Eval)} and baselines\footnote{\url{https://github.com/SVDDChallenge/CtrSVDD2024_Baseline}} are publicly accessible.
\end{abstract}

\section{Introduction}
The rapid advancement of generative artificial intelligence (AI) technologies has initiated a new era in audio deepfakes, drastically improving the quality of synthesized singing voice. Singing voice synthesis (SVS)~\cite{lu20c_interspeech, zhang2022visinger}, analogous to text-to-speech (TTS), transforms lyrics and musical scores into singing vocals. Singing voice conversion (SVC)~\cite{huang2023singing}, analogous to voice conversion (VC), transforms one singer to sound like another singer's voice without changing the lyrics and musical score.
These advancements also give rise to significant concerns within the music industry. Artists, record labels, and publishing houses are increasingly alarmed by the potential for unauthorized deepfake reproductions that closely mimic well-known singers~\cite{collins2024avoiding}, posing a direct threat to original artists' commercial value and intellectual property rights. The situation urgently calls for robust methods to protect against the unauthorized use of singing deepfake technologies.

Research has emerged towards the singing voice deepfake detection (SVDD) task as a response to synthesized singing voices. 
Our previous work introduced a multilingual in-the-wild dataset, SingFake~\cite{zang23svdd}, by collecting deepfake song clips from user-generated content websites. The label of bonafide or deepfake was identified by the uploaders and manually verified by the annotators. The synthesis methods utilized rely on uploaders to disclose and are often lacking: 60.6\% of deepfake songs in SingFake are reported with the ``Unknown'' generation method. Furthermore, out of all song clips that reported the generation method in SingFake, 92.2\% reported variants of SoVITS~\footnote{\url{https://github.com/svc-develop-team/so-vits-svc}}, indicating a potential lack of diversity.
Xie et al.~\cite{xie2023fsd} curated a controlled Chinese song dataset for fake song detection (FSD). The deepfake songs are generated using one SVS and four SVC methods applied to the bonafide songs they collected. Due to licensing restrictions, its bonafide set is not publicly available. Both works found that speech-trained deepfake detection models cannot directly work on the SVDD task, highlighting the need for a singing voice deepfake detection dataset.


In this work, we present CtrSVDD, a benchmark dataset curated for controlled SVDD with enhanced \textbf{controllability}, \textbf{diversity}, and \textbf{data openness} that we believe could further accelerate the research towards SVDD. 
Towards controllability, we manage the entire synthesis pipeline end-to-end, including specific details about the source and target datasets and the exact deepfake generation method.
Towards diversity, we include 7 SVS and 7 SVC methods to generate 188,486 (260.34 hours) deepfake song clips against 32,312 (47.64 hours) bonafide song clips for 164 singers, with an average length of 5.02 seconds. Towards data openness, our dataset is fully accessible under a CC BY-NC-ND 4.0 license. The bonafide song clips are from open-source singing datasets, while the deepfake clips include generation results from open-sourced SVS and SVC methods and those from a collaborating company, which allows us to distribute the data under the abovementioned license.



With the CtrSVDD dataset, we also present baseline systems for CtrSVDD with flexible front-end modules (encoding waveforms into feature representations) and a fixed, robust back-end module (making predictions). Using this baseline, we explored the impact of front-end features by comparing raw waveform, spectrogram-based, and cepstral coefficients (CC)-based features. The CtrSVDD dataset, baseline system implementations, and trained model weights
are publicly accessible.

\section{CtrSVDD dataset design}


The CtrSVDD dataset consists of 220,798 mono vocal clips in a total of 307.98 hours at a sample rate of 16 kHz. This section introduces the process of collecting bonafide vocal clips and generating deepfake clips. We also analyze the statistics of the resulting CtrSVDD dataset.



\subsection{Details of bonafide vocals}
\label{ssec: bonafide}

Our bonafide singing vocals are sourced from existing open singing datasets, including Mandarin singing datasets: Opencpop~\cite{wang2022opencpop}, M4Singer~\cite{zhang2022m4singer}, Kising~\cite{shi2024singing}, official ACE-Studio release~\cite{timedomain2023acesinger}, and Japanese singing datasets: Ofuton-P\footnote{\url{https://sites.google.com/view/oftn-utagoedb/\%E3\%83\%9B\%E3\%83\%BC\%E3\%83\%A0}}, Oniku Kurumi\footnote{\url{https://onikuru.info/db-download/}}, Kiritan~\cite{ogawa2021tohoku}, and JVS-MuSiC~\cite{tamaru2020jvs}. We use the official temporal segmentation in their original papers for all Mandarin datasets.
For the Japanese datasets, we performed automatic segmentation at long rests when the musical score is available or based on voice activity detection\footnote{\url{https://github.com/wiseman/py-webrtcvad}} otherwise. Following segmentation, the bona fide vocal clips have an average duration of 5.31 seconds, amounting to a total of 32,312 clips which together span 47.64 hours.

\subsection{Details of deepfake generation methods}
\label{ssec: deepfake}

We incorporate 14 deepfake systems across both SVS and SVC to cover many existing architectures, offering a comprehensive evaluation landscape. To ensure reproducibility in evaluation, we predominantly selected models from open-source toolkits, trained them on publicly available singing benchmarks, and then applied them to the bonafide singing vocals, with one exception of \textbf{A14} using a commercial system. 

\begin{figure}[]
\centering
\includegraphics[width=0.47\textwidth]{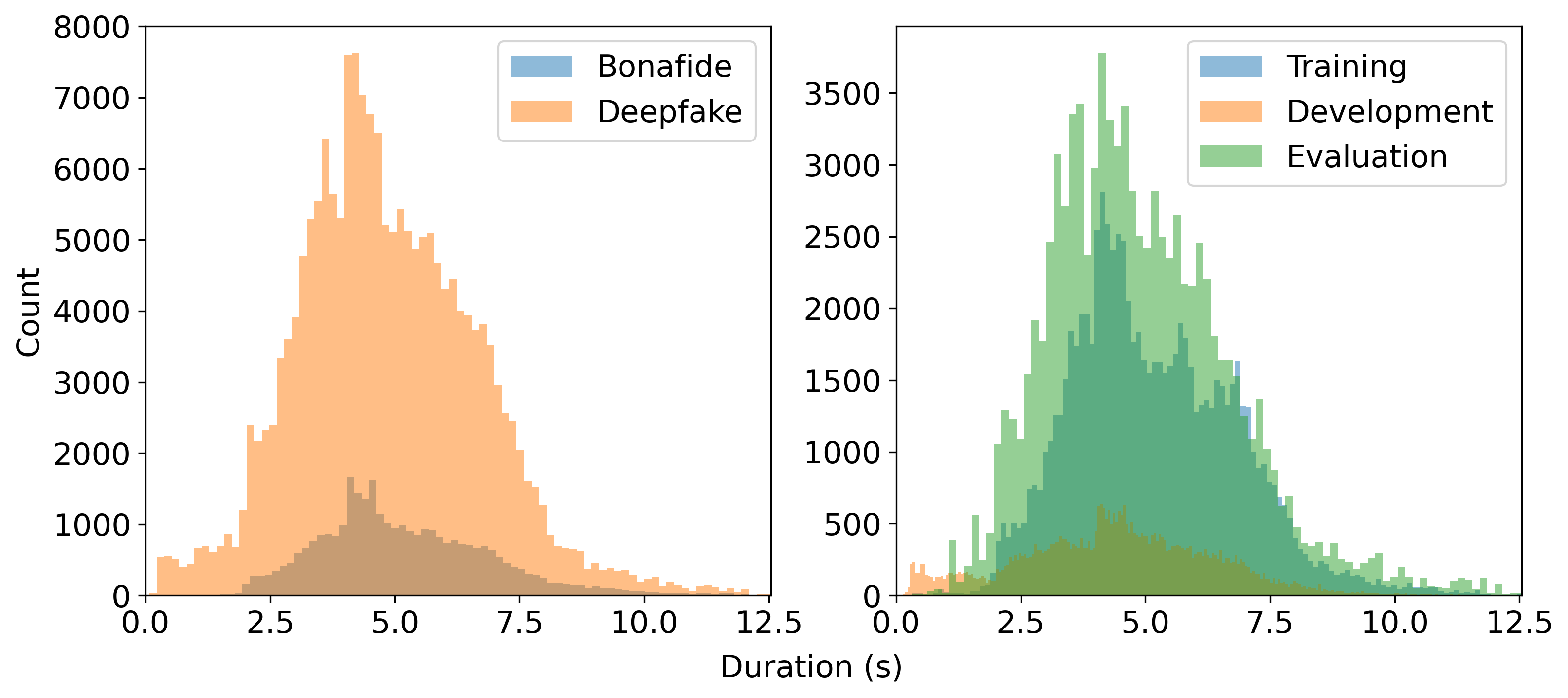}
\caption{Histogram of audio duration. 
The left subfigure shows the distribution across two classes (bonafide and deepfake), whereas the right one differentiates among the train/dev/eval splits. We exclude data exceeding three standard deviations from the mean (0.6\% of all data) for better visualization. All distributions are visualized with a 50\% opacity then overlapped for a direct comparison between them.
}
\label{fig:histogram-of-duration}
\end{figure}

\begin{table}[t]
\centering
\caption{Summary of our CtrSVDD dataset.
}
\label{table: CtrSVDD}
\resizebox{\columnwidth}{!}{
\begin{tabular}{@{}lcccc@{}}
\hline\hline
\multirow{2}{*}{Partition} & \multirow{2}{*}{\begin{tabular}[c]{@{}c@{}}\# Speakers\end{tabular}}  
& Bonafide     
& \multicolumn{2}{c}{Deepfake} \\ 
\cmidrule(l){3-5}
& &\# Utts & \# Utts   
& Attack Types     \\ 
\midrule
Train   & 59      & 12,169  & 72,235 & A01$\sim$A08     \\
Dev & 55     & ~~6,547  & 37,078 & A01$\sim$A08     \\
Eval  & 48     & 13,596  & 79,173 & A09$\sim$A14     \\ 
\hline\hline
\end{tabular}}
\end{table}

\begin{figure}[hbt!]
\centering
\includegraphics[width=0.5\textwidth]{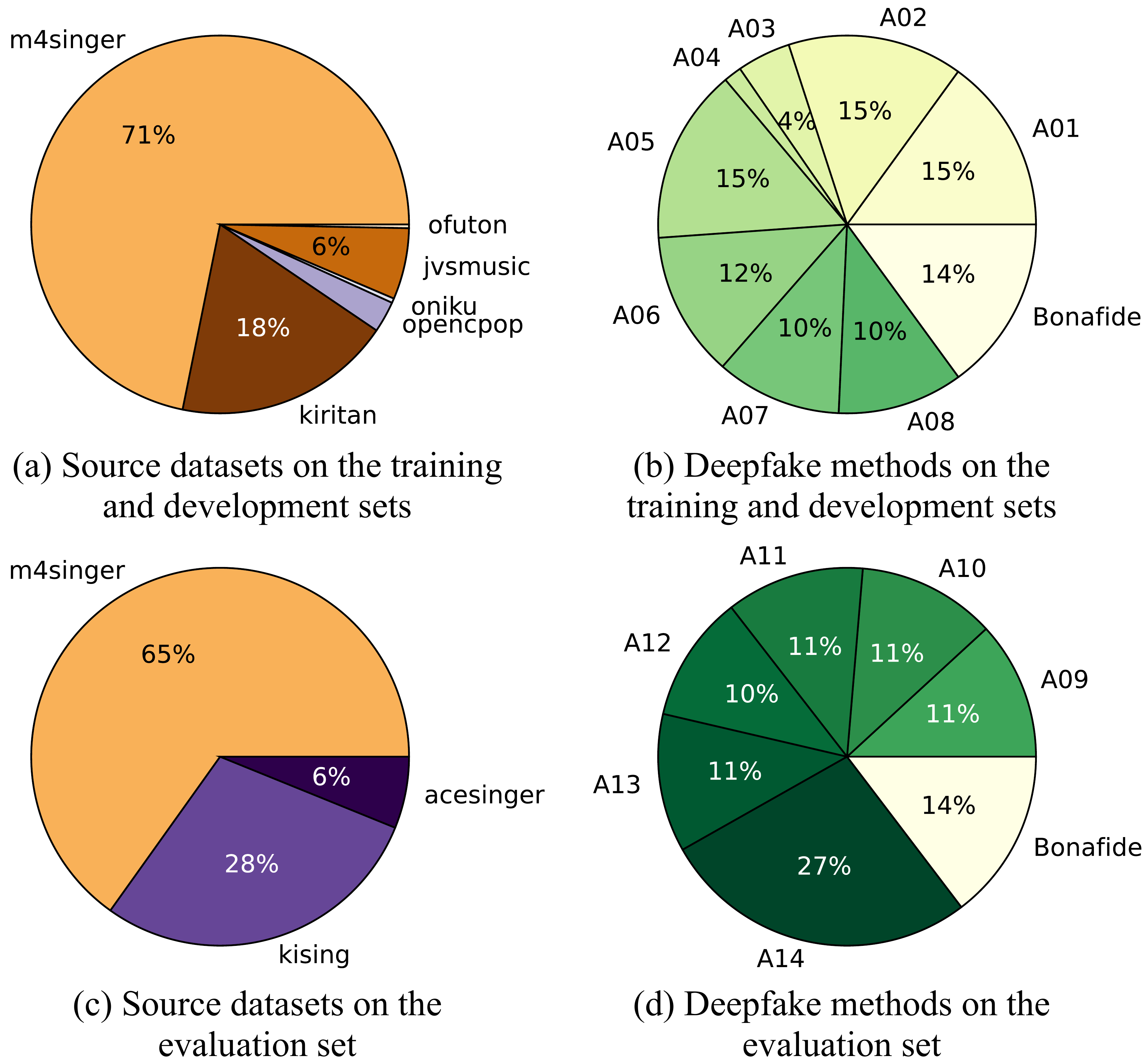}
\caption{Overview of source datasets and deepfake methods distribution on the train/dev/eval splits of our CtrSVDD data.
}
\label{fig:dataset-overview}
\end{figure}

We generate 188,486 deepfake vocal clips totaling 260.34 hours from the bonafide vocal clips, with an average length of 4.97 seconds.
Following the speech anti-spoofing benchmark dataset ASVspoof2019~\cite{wang2020asvspoof}, we use the same set of synthesis methods but different singers between training and validation sets, and hold-out singers and synthesis methods for the evaluation set. Table~\ref{table: CtrSVDD} shows the detailed summary for subsets. The audio duration distribution for song clips is shown in Figure~\ref{fig:histogram-of-duration}. 
An overview of source datasets and deepfake methods distribution is illustrated in Figure~\ref{fig:dataset-overview}. The details of each deepfake method are described as follows:

\subsubsection{Singing voice synthesis (SVS) systems}


\textbf{A01} is the non-autoregressive SVS acoustic model XiaoiceSing~\cite{lu20c_interspeech}. It employs a Transformer-based encoder-decoder architecture, similar to the FastSpeech series in the TTS domain~\cite{ren2019fastspeech, ren2020fastspeech}. The encoder and decoder are connected through a length regulator that repeats the encoder states, considering the predicted duration information. An additional HiFi-GAN vocoder~\cite{kong2020hifi} is required to synthesize the waveform output.

\textbf{A02} is the end-to-end SVS model VISinger that directly maps music scores to singing voices~\cite{zhang2022visinger}. Inspired by VITS~\cite{kim2021conditional} in TTS, VISinger employs a variational auto-encoder (VAE) architecture combined with adversarial training.

\textbf{A03} is VISinger2, an enhanced version of VISinger (A02), replacing the HiFi-GAN architecture with a differentiable digital signal processing (DDSP) vocoder~\cite{zhang23e_interspeech}.

\textbf{A04} is NNSVS~\cite{yamamoto2023nnsvs}, an open-source software that supports various neural network (NN)-based SVS systems. We select a best-performing model that combines a diffusion-based acoustic model~\cite{liu2022diffsinger} with a source-filter HiFi-GAN vocoder~\cite{yoneyama2023source}.

\textbf{A05} is Naive RNN, a non-autoregressive SVS acoustic model~\cite{shi2021sequence}. It utilizes bidirectional long-short-term memory (LSTM) layers to conduct a FastSpeech-like encoder-decoder modeling to convert music scores to spectral features. Similar to A01, an additional HiFi-GAN vocoder converts predicted spectral features to final singing voices.

\textbf{A12} is DiffSinger, which utilizes a FastSpeech backbone by adding a diffusion-based decoder (i.e., denoiser) to generate spectral features~\cite{liu2022diffsinger}. The model training is conducted in two stages, initially optimizing the FastSpeech-based SVS and then utilizing the pre-trained encoder to continue training with the diffusion-based decoder. The output spectral features from the decoder are fed into a HiFi-GAN vocoder for decoding.

\textbf{A14} is ACESinger, the singing synthesizer behind ACE-Studio~\cite{timedomain2023acesinger}. The synthesized voices are further tuned manually to remove unnatural voices as detailed in~\cite{shi2024singing}.

We utilize the training recipes in ESPnet-Muskits~\cite{shi2022muskits} to train deepfake systems of \textbf{A01}, \textbf{A02}, \textbf{A03},  \textbf{A05}, and \textbf{A12} for each database. For system \textbf{A04}, we follow the NNSVS~\cite{yamamoto2023nnsvs} corresponding recipes to optimize the system on different datasets.

\subsubsection{Singing voice conversion (SVC) systems}

\textbf{A06} is the Nagoya University (NU) SVC system~\cite{yamamoto2023comparative}, which demonstrated strong generalization capabilities and high naturalness in the SVC challenge 2023~\cite{huang2023singing}.
This model utilizes a diffusion-based acoustic model~\cite{liu2022diffsinger} and a source-filter HiFi-GAN~\cite{yoneyama2023source}. The ContentVec features~\cite{qian2022contentvec} are employed to extract the linguistic content. The model has been trained on a large-scale dataset comprising 750 hours of publicly available speech and singing data. 

\textbf{A07-A11, A13} are variations of Soft-VITS-SVC, one of the major frameworks adopted in the SVC challenge 2023~\cite{huang2023singing} by utilizng the VITS framework. The approach replaces the VITS text encoder and corresponding length regulator with pre-trained acoustic features and fundamental frequency. For \textbf{A09}, a source-filter HiFi-GAN model~\cite{yoneyama2023source} is used instead of the original HiFi-GAN model, while the source speech encoder remains the same as the original. For deepfake systems \textbf{A07}, \textbf{A08}, \textbf{A10}, \textbf{A11}, and \textbf{A13}, we employ different pre-trained acoustic representations as the prior input to the Soft-VITS-SVC system. Correspondingly, we utilize:
\begin{itemize}
    \item WavLM~\cite{chen2022wavlm} for its superior performance in SUPERB benchmark across various speech processing tasks~\cite{superb, feng2023superb},
    \item ContentVec~\cite{qian2022contentvec} as it is designed to reduce encoding of speaker information, which fits the SVC objective,
    \item MR-HuBERT~\cite{shi2023multi}, which is the first self-supervised learning framework considering multi-resolution information,
    \item WavLabLM~\cite{chen2023joint} because it has considered both noise-robustness from WavLM and multilingualism, and
    \item Chinese HuBERT~\cite{hsu2021hubert} as it is consistent with the major language (i.e., Mandarin) in the CtrSVDD bonafide singing.
\end{itemize}
For the training of Soft-VITS-SVC models, we use the Mandarin datasets (i.e., Opencpop~\cite{wang2022opencpop}, KiSing~\cite{shi2024singing, shi2022muskits}, and M4Singer~\cite{zhang2022m4singer}) and then inference each singing clip by randomly selecting another singer in the subset.


\subsection{Comparison with the FSD dataset}
The FSD dataset~\cite{xie2023fsd} is the work most similar to ours, where they utilized five deepfake systems (F01-F05) in their research. We incorporate their major systems in our CtrSVDD dataset, with A09 and A12 corresponding to their F01 and F04, respectively. Our A06 is akin to their F03, employing a diffusion-based SVC system. We opted not to include equivalents to their F02 and F05 in our dataset due to their significant resemblance to other SVC systems we have integrated, specifically A08 and A09. This selection process ensures a comprehensive yet distinct representation of various methods within our dataset, avoiding redundancy while covering a broad spectrum of techniques.

\section{Baseline systems}



Conventional hand-crafted features, such as linear frequency cepstral coefficients (LFCC), have shown promising results in speech deepfake detection. Moreover, recent advancements in end-to-end learning approaches, such as raw waveform-based models~\cite{Jung2021AASIST}, have demonstrated competitive performance. However, the effectiveness of these features in the context of singing voice deepfake detection remains largely unexplored. Therefore, we propose to systematically evaluate a diverse range of representations to gain insights into their effectiveness and robustness in detecting singing voice deepfakes.

To this end, we design a versatile baseline framework to facilitate a fair evaluation of diverse front-end representations. The system first extracts features from interchangeable front-end modules (Section \ref{ssec:frontends}), then employs downsampling residual blocks, followed by a graph attention module (Section \ref{ssec:backend}) to aggregate spatial and temporal information. Finally, an output layer produces a probability score reflecting the deepfake likelihood of each song clip.

\begin{table*}[hbt!]
\caption{Evaluation results of baseline systems on the evaluation set. Best performing results for each category are illustrated in bold.}
\label{table: results}
\begin{tabular}{c|c|cccccc}
\hline
\hline
\multirow{2}{*}{Frontend} & \multirow{2}{*}{EER (\%)} & \multicolumn{6}{c}{Per-method EER (\%)}                                                                                                   \\ 
                          &                           & \textbf{A09}         & \textbf{A10}         & \textbf{A11}         & \textbf{A12}          & \textbf{A13}         & \textbf{A14}          \\ \hline
Spectrogram               & 25.50 ± 0.09              & 32.02 ± 0.09         & 14.03 ± 0.10         & 14.67 ± 0.08         & 35.18 ± 0.31          & 18.10 ± 0.10         & 28.55 ± 0.17          \\
Mel-Spectrogram           & 25.19 ± 0.10              & 25.29 ± 0.14         & 15.95 ± 0.15         & 37.31 ± 0.09         & 29.28 ± 0.25          & 12.86 ± 0.08         & 27.54 ± 0.11          \\
MFCC                      & 26.67 ± 0.07              & 6.87 ± 0.10          & 2.50 ± 0.07          & 4.18 ± 0.06          & 45.57 ± 0.11          & 3.28 ± 0.04          & 42.98 ± 0.08          \\ 
LFCC                      & 16.15 ± 0.06              & \textbf{5.35 ± 0.07} & 2.92 ± 0.04          & 5.84 ± 0.07          & 29.47 ± 0.06          & 3.65 ± 0.05          & 24.00 ± 0.10          \\
Raw Waveform              & \textbf{13.75 ± 0.11}     & 6.72 ± 0.06          & \textbf{0.96 ± 0.05} & \textbf{3.59 ± 0.06} & \textbf{26.83 ± 0.10} & \textbf{0.95 ± 0.04} & \textbf{19.03 ± 0.12} \\
\hline \hline
\end{tabular}
\end{table*}

\subsection{Frontends}
\label{ssec:frontends}
The front end refers to the pre-processing part of the network that converts raw audio samples into features, which the backend neural network can use to make predictions.

\textbf{Spectrogram}. We employ a normalized power spectrogram with a 512-sample window and hop size of 160 samples. 

\textbf{Mel-Spectrogram}. We employ a mel-filterbank with 80 bands on the spectrogram.

\textbf{Mel-Frequency Cepstral Coefficients (MFCC)}. We extract 40 MFCC bands with spectral processing parameters similar to those used in LFCC.

\textbf{Linear Frequency Cepstral Coefficients (LFCC)}. We employ 20 filters from 0 to 8 kHz to extract 60 coefficients from the audio signal, with type-II discrete Fourier transform (DCT) and ortho-norm for normalization, with a window length of 512 samples, and a hop length of 160 samples.

\textbf{Raw waveform}. We follow~\cite{Jung2021AASIST} to employ a RawNet2-style~\cite{jung2020improved} learnable SincConv layer with 70 filters.

The residual blocks following front-end modules are implemented as two sequential batch normalization, SELU activation, and convolution blocks, with a residual connection between each block's input and output; max pooling is applied before outputting. The first set of batch normalization and activation is dropped for the first residual block. We employ four residual blocks for all spectral and cepstral features; for the raw waveform feature, we employ six residual blocks to match~\cite{Jung2021AASIST}. The first two residual blocks have 32 filters, and the remaining ones have 64 filters. A linear layer is then used to connect it to the backend.

\subsection{Backend}
\label{ssec:backend}
We follow~\cite{Jung2021AASIST} in our backend implementation, which consists of fully connected graph attention networks for spectral and temporal domains, then combined into a spatial-temporal graph and processed using heterogeneous stacking graph attention layers and four graph pooling layers. We connect the readout from graph pooling to a single neuron output with a linear layer. The logits output by the network (before activation) indicates the likelihood that a given song clip is bonafide.

\section{Experiments and results} 

\subsection{Experimental setup}
We formulate the SVDD task as a binary classification task in alignment with the methodologies proposed by~\cite{zang23svdd, xie2023fsd}. SVDD models assign a continuous score to each vocal clip, with higher values indicating authentic singing and lower values suggesting deepfake ones. while a threshold is needed for model deployment in practice, using such a threshold may introduce unnecessary bias for model comparison. Instead, 
we employ the Equal Error Rate (EER) as the evaluation metric, which denotes the point at which the rates of false acceptances and false rejections are equal. This metric, distinct from accuracy, is not influenced by the choice of threshold, making it particularly apt for evaluating the performance of SVDD systems. A lower EER is indicative of a system's superior performance.



We consistently apply a fixed random seed across all systems, utilizing the Adam optimizer, a batch size of 24, a learning rate of \texttt{1e-3}, and a weight decay of \texttt{1e-9}. Additionally, we employ a cosine annealing learning rate schedule that cycles to \texttt{1e-6} every 10 epochs.
We use binary focal loss~\cite{lin2017focal}, a generalized version of the binary cross-entropy loss, with focusing parameter ($\gamma$) as 2 and positive example weight ($\alpha$) as 0.25. To ensure uniformity in input length, each song clip is either randomly cropped or extended to 4 seconds for batch formation during training, validation, and evaluation phases. Every system is trained for 100 epochs, after which the model checkpoint with the lowest validation EER is selected for evaluation.

In evaluation, we apply 5 different random seeds to trim vocal clips, creating 5 variations of the test set. We report the mean and standard deviation of the EER across these versions to assess model robustness against random time shifts.
All experiments are performed on a single RTX 4090 GPU. The training time for the raw-waveform-based model is slightly longer than 24 hours, while the spectrogram-based model trains for around 7 hours and all other frontend features for about 5 hours.

\subsection{Results and discussions}
Table~\ref{table: results} presents the system performance results. The small standard deviation observed across all EERs suggests consistent and stable predictions across random window shifts within each song clip, lending statistical significance to comparisons among the baseline systems.

\textbf{Overall EERs.} Amongst all frontends for our baseline systems, the raw-waveform-based system achieves the lowest overall EER, closely succeeded by the LFCC-based system. Systems based on spectrograms, mel-spectrograms, and MFCCs exhibit comparable overall performances, trailing behind raw-waveform-based and LFCC-based systems by a large margin. 

\textbf{Per-method EERs.} We observe that the performance gap between the top two performing systems and the remaining methods is notably large in \textbf{A09-A11} and \textbf{A13}, which are variations of the Soft-VITS-SVC with different text encoders. This suggests that the top-performing frontends generalize against unseen content encoding methods better. Conversely, all systems perform much weaker for \textbf{A12} and \textbf{A14}. \textbf{A14}, as a commercial black-box system, has an undisclosed architecture, whereas \textbf{A12} employs a diffusion-based decoder on top of the FastSpeech backbone, which is distinct from other deepfake generation approaches.
We speculate that the divergence of these methods from the training distribution might prevent SVDD systems from effectively distinguishing them from bonafide singing, indicating a challenge in learning discriminative representations for these unique deepfake techniques.

\begin{figure}[]
\centering
\includegraphics[width=0.46\textwidth]{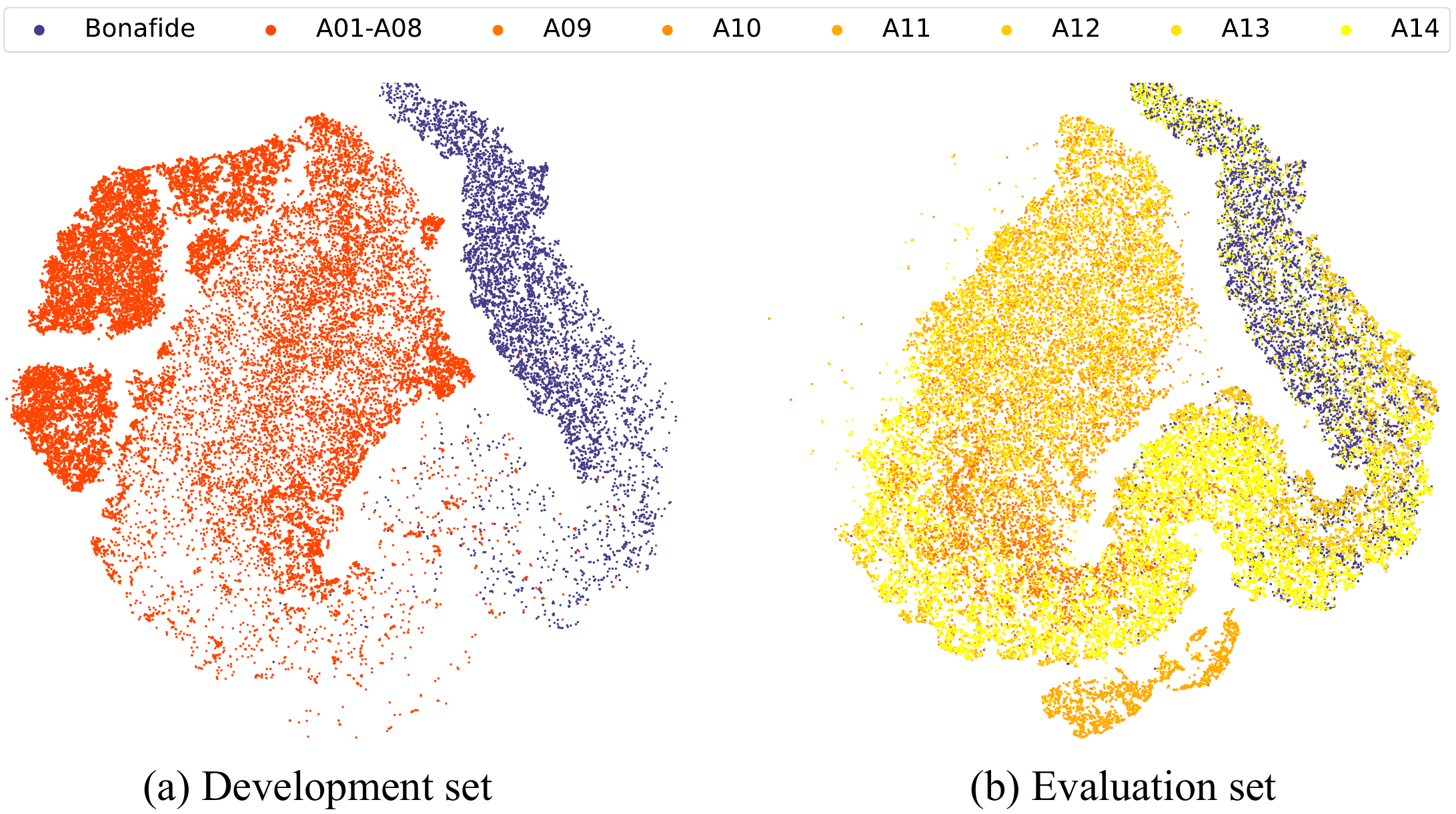}
\caption{t-SNE visualization of the learned representation for the raw-waveform-based baseline system on both development and evaluation sets.
Best viewed in color.}
\label{fig:tsne}
\end{figure}

To test this hypothesis, we visualize the learned representation before the final linear layer of the raw-waveform frontend using t-SNE~\cite{van2008visualizing} in Figure~\ref{fig:tsne}. The visualizations on the development and evaluation sets use the same coordinate system.

As depicted in Figure~\ref{fig:tsne}, the distribution of bonafide singing remains consistent across both datasets. However, the distributions of deepfake singing, especially for methods \textbf{A12} and \textbf{A14}, exhibit significant overlap with that of bonafide singing. This overlap explains the reduced performance observed with these methods. It suggests that although top-performing systems can differentiate between deepfake and bonafide singing when the deepfake characteristics are similar to those encountered during training (\textbf{A09-A11} and \textbf{A13}), they struggle to accurately represent deepfake methods that deviate further from the training distribution. This underscores the need for research on SVDD systems with improved generalization ability to unseen deepfake techniques, as these techniques are rapidly advancing.

\section{Conclusion}
We present CtrSVDD, a dataset for controlled singing voice deepfake detection. CtrSVDD addresses key limitations in existing SVDD datasets by providing enhanced controllability, diversity, and data openness, comprising a large-scale collection of 220,798 vocal clips totaling 307.98 hours. To facilitate SVDD research using CtrSVDD, we also presented a versatile baseline system that allows for interchangeable front-end feature extraction modules. Our experiments demonstrated the importance of feature selection, with raw waveform and LFCC front-ends exhibiting the most robust performance. However, the results also highlighted a lack of generalization towards unseen deepfake methods, underscoring the need for more generalizable SVDD systems. By releasing CtrSVDD, baseline implementations, and pre-trained model weights, we aim to accelerate research for the SVDD task.

\vfill\pagebreak

\section{Acknowledgments}
This work is supported in part by a New York State Center of Excellence in Data Science award, National Institute of Justice (NIJ) Graduate Research Fellowship Award 15PNIJ-23-GG-01933-RESS, National Science Foundation (NSF) grants 1846184 and 2222129, synergistic activities funded by NSF grant DGE-1922591, and JST CREST JPMJCR19A3, Japan.




\bibliographystyle{IEEEtran}
\bibliography{mybib}

\end{document}